\documentclass[aps,pra,twocolumn,showpacs,superscriptaddress,floatfix]{revtex4-1}
\usepackage{graphicx,amsmath,amssymb}
\usepackage[usenames]{color}
\usepackage[dvipsnames]{xcolor}
\usepackage[colorlinks=true,linkcolor=blue,anchorcolor=blue,citecolor=blue,urlcolor=blue]{hyperref}
\begin{document}

\title{Globalized critical quantum metrology in dynamics of quantum Rabi model by auxiliary nonlinear term}

\author{Qiu-Yi Chen}
\affiliation{School of Physical Science and Technology, Lanzhou University, Lanzhou 730000, China}
\affiliation{Key Laboratory for Quantum Theory and Applications of MoE, Lanzhou Center for Theoretical Physics, Lanzhou University, Lanzhou 730000, China}

\author{Feng Qiao}
\affiliation{School of Physical Science and Technology, Lanzhou University, Lanzhou 730000, China}
\affiliation{Key Laboratory for Quantum Theory and Applications of MoE, Lanzhou Center for Theoretical Physics, Lanzhou University, Lanzhou 730000, China}

\author{Zu-Jian Ying}
\email{yingzj@lzu.edu.cn}
\affiliation{School of Physical Science and Technology, Lanzhou University, Lanzhou 730000, China}
\affiliation{Key Laboratory for Quantum Theory and Applications of MoE, Lanzhou Center for Theoretical Physics, Lanzhou University, Lanzhou 730000, China}

\begin{abstract}
Quantum Rabi model (QRM) is a fundamental model for light-matter interactions, the finite-component quantum phase transition (QPT) in the QRM has established a paradigmatic application for critical quantum metrology (CQM). However, such a paradigmatic application is restricted to a local regime of the QPT which has only a single critical point. In this work we propose a globalized CQM in the QRM by introducing an auxiliary nonlinear term which is realizable and can extend the critical point to a continuous critical regime. As a consequence, a high measurement precision is globally available over the entire coupling regime from the original critical point of the QRM down to the weak-coupling limit, as demonstrated by the globally accessible diverging quantum Fisher information in dynamics.  We illustrate a measurement scheme by quadrature dynamics, with globally criticality-enhanced inverted variance as well as the scaling relation with respect to finite frequencies. In particular, we find that the globally high measurement precisions still survive in the presence of decoherence. Our proposal paves a way to break the local limitation of QPT of the QRM in CQM and enables a broader application, with implications of applicability in realistic situation.
\end{abstract}
\pacs{ }
\maketitle


\section{Introduction}

During the past decades the continuous experimental efforts in coupling enhancements for light-matter interactions have built up an ideal platform for the explorations of advanced quantum technologies, with the merits of high controllability and tunability~\cite{Diaz2019RevModPhy,Kockum2019NRP,PRX-Xie-Anistropy,Eckle-Book-Models,JC-Larson2021,
Boite2020,Qin-ExpLightMatter-2018,WangYouJQ2023DeepStrong,Qin2024PhysRep,LiPengBo-Magnon-PRL-2024}.
The finite-component quantum phase
transitions in light-matter interactions (QPTs) have found potential applications in critical quantum metrology (CQM)~\cite{Garbe2020,Montenegro2021-Metrology,Chu2021-Metrology,Garbe2021-Metrology,Ilias2022-Metrology, Ying2022-Metrology,YangZheng2023SciChina,Gietka2023PRL-Squeezing,Hotter2024-Metrology,Alushi2024PRL,Mukhopadhyay2024PRL,Mihailescuy2024,
Ying-Topo-JC-nonHermitian-Fisher,*Ying-Topo-JC-nonHermitian-Fisher-Cover,Ying-g2hz-QFI-2024,*Ying-g2hz-QFI-2024-Cover,Ying-g1g2hz-QFI-2025,Ying2025g2A4,*Ying2025g2A4-Cover,Ying-g2Stark-QFI-2025}.

In fact, light-matter interactions have a wide relevance in quantum information and quantum
computation~\cite{Diaz2019RevModPhy,Romero2012,Stassi2020QuComput,Stassi2018,Macri2018},
quantum metrology~\cite{Garbe2020,Montenegro2021-Metrology,Chu2021-Metrology,Garbe2021-Metrology,Ilias2022-Metrology, Ying2022-Metrology,Gietka2023PRL-Squeezing,YangZheng2023SciChina,Hotter2024-Metrology,Alushi2024PRL,Mukhopadhyay2024PRL,Mihailescuy2024,
Ying-Topo-JC-nonHermitian-Fisher,*Ying-Topo-JC-nonHermitian-Fisher-Cover,Ying-g2hz-QFI-2024,*Ying-g2hz-QFI-2024-Cover,Ying-g1g2hz-QFI-2025,Ying2025g2A4,*Ying2025g2A4-Cover,Ying-g2Stark-QFI-2025},
condensed matter~\cite{Kockum2019NRP}, 
and cold atoms~\cite{LinRashbaBECExp2013Review,LinRashbaBECExp2011,LiuYing02025exoticSOC2Ring,*LiuYing02025exoticSOC2Ring-Cover,LiuYing02025KaleidoscopeDDI}. A most fundamental model of light-matter interactions is the quantum Rabi model (QRM)~\cite{rabi1936,Braak2011,Rabi-Braak,Eckle-Book-Models}. With the entrance into the era of  ultra-strong~\cite{Diaz2019RevModPhy,Kockum2019NRP,Qin2024PhysRep} and even deep-strong~\cite{WangYouJQ2023DeepStrong,Yoshihara2017NatPhys,Bayer2017DeepStrong} couplings, a fascinating phenomenon that the QRM and its extensions manifest is the existence of finite-component quantum phase transitions (QPTs)~\cite{Ashhab2013,Ying2015,Liu2021AQT,Hwang2015PRL,Hwang2016PRL,Irish2017,
Ying-g2hz-QFI-2024,*Ying-g2hz-QFI-2024-Cover,Ying-g1g2hz-QFI-2025,Ying2025g2A4,*Ying2025g2A4-Cover,Ying-g2Stark-QFI-2025,
LiuM2017PRL,Ying-2018-arxiv,Ying2020-nonlinear-bias,Ying-2021-AQT,*Ying-2021-AQT-Cover,
Ying-gapped-top,
Ying-Stark-top,*Ying-Stark-top-Cover,
Ying-Spin-Winding,*Ying-Spin-Winding-Cover,
Ying-JC-winding,Ying-Topo-JC-nonHermitian,*Ying-Topo-JC-nonHermitian-Cover,Ying-Topo-JC-nonHermitian-Fisher,*Ying-Topo-JC-nonHermitian-Fisher-Cover,Ying-gC-by-QFI-2024, Grimaudo2022q2QPT,Grimaudo2023-Entropy,Grimaudo2024PRR,Zhu2024PRL,DeepStrong-JC-Huang-2024,PengJie2019,Padilla2022,Gao2022Rabi-dimer,GaoXL2025SPT}. Although the thermodynamical limit for QPTs in condense matter is absent in the QRM which is actually a finite-component system, the low-frequency limit is a replacement for the occurrence of the finite-component QPT~\cite{Ashhab2013,Ying2015,Hwang2015PRL,LiuM2017PRL,Irish2017}. Indeed,
the QPT in the QRM has established a paradigmatic application for CQM~\cite{Garbe2020,Chu2021-Metrology,Garbe2021-Metrology,Ilias2022-Metrology, Ying2022-Metrology,Gietka2023PRL-Squeezing,Hotter2024-Metrology,Alushi2024PRL,Mukhopadhyay2024PRL,Mihailescuy2024,
Ying-Topo-JC-nonHermitian-Fisher,*Ying-Topo-JC-nonHermitian-Fisher-Cover,Ying-g2hz-QFI-2024,*Ying-g2hz-QFI-2024-Cover,Ying-g1g2hz-QFI-2025,Ying2025g2A4,*Ying2025g2A4-Cover}.
However, as the QPT of the QRM has only a single critical point, such a paradigmatic application is restricted to a local coupling regime of the QPT. In such a situation, globalization of CQM to wider measurement parameter regime is desirable.

The standard QRM involves a linear coupling by processes of one-photon abortion and emission, while nonlinear coupling usually involves two-photon processes~\cite{Felicetti2018-mixed-TPP-SPP,Felicetti2015-TwoPhotonProcess,e-collpase-Garbe-2017,Rico2020,e-collpase-Duan-2016,CongLei2019,Ying-2018-arxiv,Ying2020-nonlinear-bias}. Efforts to globalize the CQM have been exerted on nonlinear couplings by quantum resource combinations of the two-photon coupling with asymmetry and bias field~\cite{Ying-g2hz-QFI-2024,*Ying-g2hz-QFI-2024-Cover}, one-photon coupling~\cite{Ying-g1g2hz-QFI-2025}, Stark coupling~\cite{Ying-g2Stark-QFI-2025} or quartic term~\cite{Ying2025g2A4,*Ying2025g2A4-Cover}. However, few globalization of CQM has been proposed for the linear coupling in the standard QRM which is actually more original and paradigmatic~\cite{Garbe2020}. Combining with two-photon coupling and bias can extend the regime of CQM for the linear coupling in the QRM~\cite{Ying2022-Metrology}, which resorts to an emerging first-order-like QPT, while globalization of CQM by maintaining the original and well-established second-order QPT is still in need.

In the present work we propose to globalize the CQM in the QRM by adding an auxiliary nonlinear term which can extend the critical point to a continuous critical regime. In such a protocol, as demonstrated by the globally accessible quantum Fisher information (QFI) in dynamics, a high measurement precision is globally available over the entire coupling regime from the original critical point of the QRM down to the weak-coupling limit.  We further provide a measurement scheme by quadrature measurements, with the global precision illustrated  by the inverted variance and the scaling relation with respect to finite frequencies. Finally we show that the globally enhanced measurement precision still survive in the presence of decoherence. Our proposal breaks the local limitation of QPT of the QRM and paves a way for a broader application in CQM.

The paper is organized as follows.
Section~\ref{Section-Model}
introduces the QRM with auxiliary nonlinear term and derives the effective low-energy Hamiltonian. Critical point is extracted.
Section~\ref{Section-dynamics-QFI}
derives and addresses the QFI in dynamics.
Section~\ref{Sect-global-CQM}
demonstrates globalized critical quantum metrology with divergent QFI available in entire coupling regime. 
Section~\ref{Sect-Measurement-schemes}
describes the measurement scheme by quadrature dynamics. The inverted variance is shown to scale with the QFI.
Section~\ref{Sect-Scaling-relation}
estimates the discrepancy between finite frequencies and low-frequency limit. Scaling relation is obtained for the the discrepancy. 
Section~\ref{Sect-decoherence}
discusses the surviving measurement precision in the presence of decoherence.
Finally, Section~\ref{Section-Conclusion}
summarizes our conclusions.

\section{Model and regulated transition point}\label{Section-Model}

The standard QRM in light-matter interactions~\cite{rabi1936,Braak2011,Rabi-Braak,Eckle-Book-Models} is described by the Hamiltonian
\begin{equation}\label{equ1}
	H_{Rabi}=\omega a^\dag a+\frac{\Omega}{2}\sigma_{z}+\frac{\sqrt{\omega\Omega}}{2}g(a+a^\dag)\sigma_{x},
\end{equation}
where $\Omega$ is the frequency of the two-level system, $\sigma_{x,y,z}$ are the Pauli operators, $a^\dag(a)$ is the bosonic creation (annihilation) operator with the frequency $\omega$, and $g$ is the normalized coupling strength. In the low-frequency limit the QPT occurs at
\begin{equation}\label{eq-gc}
	g=g_c^{(0)}=1,
\end{equation}
which is however only a single point.
To globalize the CQM of the QRM we should enable the regulating of the phase transition point, for this sake we add a neutral quadratic term auxiliary to the QRM:
\begin{equation}\label{equ2}
	H=\omega a^\dag a+\frac{\Omega}{2}\sigma_{z}+\frac{\sqrt{\omega\Omega}}{2}g(a+a^\dag)\sigma_{x}+\lambda(a+a^\dag)^2.
\end{equation}
This quadratic term is neutral in the sense that it is the same for different spin components, unlike the coupling term.
Such a quadratic term can be realized and adjustable in hybrid systems with optomechanical coupling~\cite{Chen2021A2NC,Lv2018A2PRA,Lv2018A2PhysRevApplied} or Kerr magnons~\cite{LiuGang2023}. 

By the squeezing transformation with the squeezing parameter $r=\frac{1}{4}\ln(1+\frac{4\lambda}{\omega})$, the Hamiltonian can be written as
\begin{equation}\label{equ3}
\begin{aligned}
	H_{r} & =e^{-\frac{r}{2}(a^2-a^{\dag2})}He^{\frac{r}{2}(a^2-a^{\dag2})} \\
		& =\omega \left( 1+\frac{4\lambda}{\omega}\right) ^\frac{1}{2}a^\dag a+\frac{\Omega}{2}\sigma_{z}\\
		&\quad+\frac{\sqrt{\omega\Omega}}{2}g\left( 1+\frac{4\lambda}{\omega}\right) ^{-\frac{1}{4}}(a+a^\dag)\sigma_{x}.
\end{aligned}
\end{equation}

In the case of the low-frequency limit $\omega\ll\Omega$ and $\frac{\omega g^2}{\omega+4\lambda}<1$, we perform the Schrieffer-Wolff transform on the Hamiltonian
\begin{equation}\label{equ4}
\begin{aligned}
	H_{S} & =e^{-S}H_{r}e^{S} \\
		& =\omega \left( 1+\frac{4\lambda}{\omega}\right) ^\frac{1}{2}a^\dag 	a+\frac{\Omega}{2}\sigma_{z}\\
		&\quad-\frac{\omega g^2}{4}\left( 1+\frac{4\lambda}{\omega}\right) ^\frac{3}{2}(a+a^\dag)^2\sigma_{z}  +\mathcal{O}\left(\frac{\omega^2}{\Omega} \right) ,
\end{aligned}
\end{equation}
where
\begin{equation}\label{equ5}
	\begin{aligned}
	S=&-\frac{ig}{2}\sqrt{\frac{\omega}{\Omega}}\left( 1+\frac{4\lambda}{\omega}\right) ^{-\frac{1}{4}}(a+a^\dag)\sigma_{y}\\
	&+\frac{ig^3}{6}\left( \frac{\omega}{\Omega}\right)^\frac{3}{2}\left( 1+\frac{4\lambda}{\omega}\right) ^{-\frac{3}{4}}(a+a^\dag)^3\sigma_{y}\\
	&+\mathcal{O}\left[ \left(\frac{\omega}{\Omega}\right)^\frac{3}{2}\right].
\end{aligned}
\end{equation}

An effective low-energy Hamiltonian is obtained  by projecting  Eq.\eqref{equ4} onto the spin-down subspace
\begin{equation}
	H_{np}=\frac{\sqrt{\omega^2+4\lambda\omega}}{2}\left[P^2+\epsilon_gX^2\right], \label{Hnp}
\end{equation}
where $\epsilon_g=\left(1-\frac{\omega g^2}{\omega+4\lambda}\right)$, the quadrature operators are defined as $X=(a+a^\dagger)/\sqrt{2}$ and $P=i(a^\dagger-a)/\sqrt{2}$. From this effective model \eqref{Hnp}, we see that the phase transition point moves to
\begin{equation}\label{eq-gC}
	g=g_c^{(\lambda)}=\sqrt{1+4\lambda/\omega},
\end{equation}
which can be continuously regulated by $\lambda$. Indeed, via adjusting the auxiliary nonlinear term by
\begin{equation}
	\lambda=\lambda_c=(1-g^2)\frac{\omega}{4},  \label{eq-A2C}
\end{equation}
one can tune the critical point to a needed coupling $g$, which enables the globalization of CQM as addressed in the next sections.

\section{The QFI of critical quantum dynamics}\label{Section-dynamics-QFI}

The performance of quantum sensing depends on the sensitivity to distinguish between states $\lvert\psi_{\zeta}\rangle$ and $\lvert\psi_{\zeta+\delta\zeta}\rangle$ of nearby parameters, which is acutally the susceptibility of fidelity equivalent to the QFI~\cite{Gu-FidelityQPT-2010,You-FidelityQPT-2007,You-FidelityQPT-2015,Ying-g1g2hz-QFI-2025,Zhou-FidelityQPT-2008}. Indeed, as limited by the quantum
Cram\'{e}r-Rao theorem~\cite{Cramer-Rao-bound} the measurement precision of experimental estimation on
a parameter $\zeta $ is bounded by $F_\zeta^{1/2}$, where $F_{\zeta}$ is the QFI
defined as \cite{Cramer-Rao-bound,Taddei2013FisherInfo,RamsPRX2018}
\begin{equation}
F_{\zeta} =4\left[ \langle \partial_\zeta \psi_\zeta | \partial_\zeta \psi_\zeta  \rangle
-\left\vert \langle
\partial_\zeta \psi_\zeta  |\psi_\zeta \rangle
\right\vert ^{2}\right]   \label{Fq}
\end{equation}%
for a pure state $|\psi_\zeta \rangle $. For a non-degenerate eigenstate of a real Hamiltonian, as it is for most ground states in light-matter interactions, the QFI can be simplified to be $F_{\zeta} =4\langle \partial_\zeta \psi_\zeta | \partial_\zeta \psi_\zeta  \rangle$~\cite{Ying-gC-by-QFI-2024}. In quantum metrology, a larger QFI means a higher measurement precision available.

We consider the Hamiltonian about the parameter $\zeta$, $H_\zeta=H_0+\zeta H_1$, whose dynamical properties are determined by unitary evolution $U_\zeta=\exp(-iH_\zeta t)$. In such a situation the QFI can be written as
\begin{equation}
	F_\zeta=4Var[h_\zeta]_{\lvert\varphi\rangle},
\end{equation}
where $Var[...]_{\lvert\varphi\rangle}$ represents the variance corresponding to the initial state $\lvert\varphi\rangle$, and $h_\zeta=iU_{\zeta}^\dagger(\partial_\zeta U_\zeta)$. And $h_\zeta$ can be transformed into the following form~\cite{Pang2017FqHt}
\begin{equation}\label{equ7}
	\begin{aligned}
	h_\zeta&=\int_{0}^t e^{iH_\zeta s}H_1 e^{-iH_\zeta s}ds\\
	&=-i\sum_{n=0}^{\infty}\frac{(it)^{n+1}}{(n+1)!}[H_\zeta, H_1]_n ,
\end{aligned}
\end{equation}
where $[H_\zeta, H_1]_{n+1}$=$[H_\zeta, [H_\zeta, H_1]_n]$, and $[H_\zeta, H_1]_0=H_1$.

In order to simplify $h_\zeta$, we require the Hamiltonian to satisfy a special reciprocal relation~\cite{Chu2021-Metrology, Pang2014PRAFqGeneralH}
\begin{equation}\label{equ8}
	[H_\zeta,\Lambda]=\sqrt{\epsilon}\Lambda,
\end{equation}
where $\epsilon$ is dependent on the parameter $\zeta$, and $\Lambda=i\sqrt{\epsilon}M-N$ with $M=-i[H_0, H_1]$, $N=-[H_\zeta,[H_0, H_1]]$. In this case, $[H_\zeta, H_1]_n$ can be split into two parts
\begin{equation}\label{equ9}
	\begin{aligned}
	&[H_\zeta, H_1]_{2n+1}=i\epsilon^nM,\\
	&[H_\zeta, H_1]_{2n+2}=-\epsilon^nN.
	\end{aligned}
\end{equation}
Therefore, simplified Eq.\eqref{equ7} can be obtained
\begin{equation}\label{equ10}
	h_\zeta=H_1t+\frac{\cos(\sqrt{\epsilon}t)-1}{\epsilon}M-\frac{\sin(\sqrt{\epsilon}t)-\sqrt{\epsilon}t}{\epsilon\sqrt{\epsilon}}N.
\end{equation}
The energy gap closes near the point of phase transition resulting in $\sqrt{\epsilon}\rightarrow 0$. Under the constraint of condition $\sqrt{\epsilon}t\sim\mathcal{O}(1)$, the approximate value of QFI can be obtained by selecting the largest divergence scale $\epsilon^{-3}$ term in $4Var[h_\zeta]_{\lvert\varphi\rangle}$
\begin{equation}\label{equ11}
	F_\zeta\simeq 4\frac{[\sin(\sqrt{\epsilon}t)-\sqrt{\epsilon}t]^2}{\epsilon^3}Var[N]_{\lvert\varphi\rangle}.
\end{equation}

We apply this approach to Eq.~\eqref{Hnp} by choosing $\zeta=\epsilon_g$ and obtain the QFI about the parameter $g$
	\begin{equation}\label{Eq-Fg}
	\begin{aligned}
		F_g&=(\partial_g\zeta)^2F_\zeta \\
		& \simeq 16\left( \frac{\omega g}{\omega+4\lambda}\right) ^2\frac{[\sin(\sqrt{\epsilon} t)-\sqrt{\epsilon} t]^2}{\epsilon^3}Var[N]_{\lvert\varphi\rangle}
	\end{aligned},
\end{equation}
where $\epsilon=4\omega(\omega+4\lambda)\epsilon_g$, and $N=(\omega^2+4\lambda\omega)^\frac{3}{2}(P^2-\epsilon_gX^2)$. Here the results are derived for the regime $g<g_c^{{\lambda}}$, while the expressions in the regime $g>g_c^{{\lambda}}$ can be found in Appendix~\ref{Appendix-Beyond-gc}.

\section{Globalized critical quantum metrology}\label{Sect-global-CQM}

For CQM we need to check whether we have criticality-enhanced measurement precision. For quantum metrology, the square root of the QFI is the upper bound of measurement precision in quantum metrology, as established by the quantum
Cram\'{e}r-Rao theorem~\cite{Cramer-Rao-bound}. In this section we will see divergent QFI in dynamics arising from the criticality and the divergent QFI can be globally acquired.

\subsection{QFI enhanced by critical quantum dynamics}

\begin{figure}[htbp]
	\centering
	\includegraphics[width=7cm]{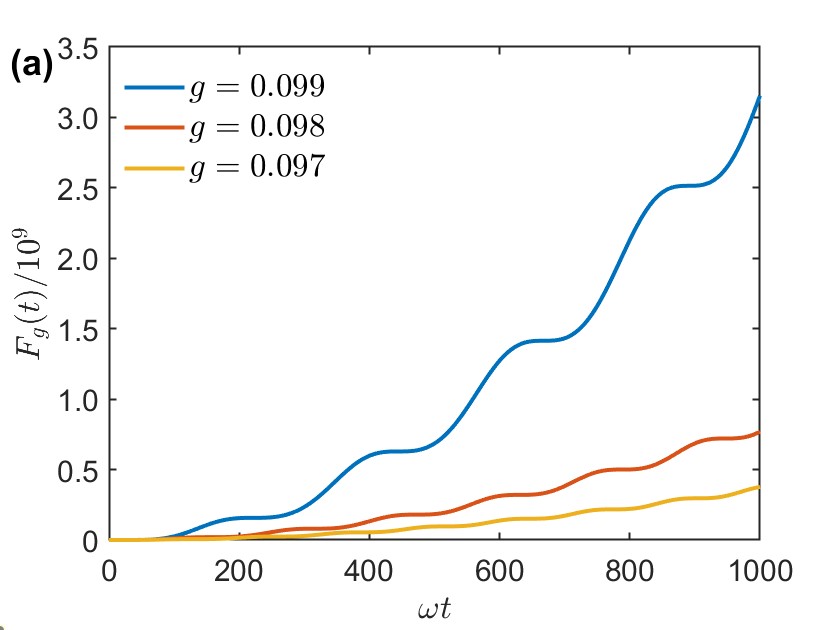}
	\includegraphics[width=7cm]{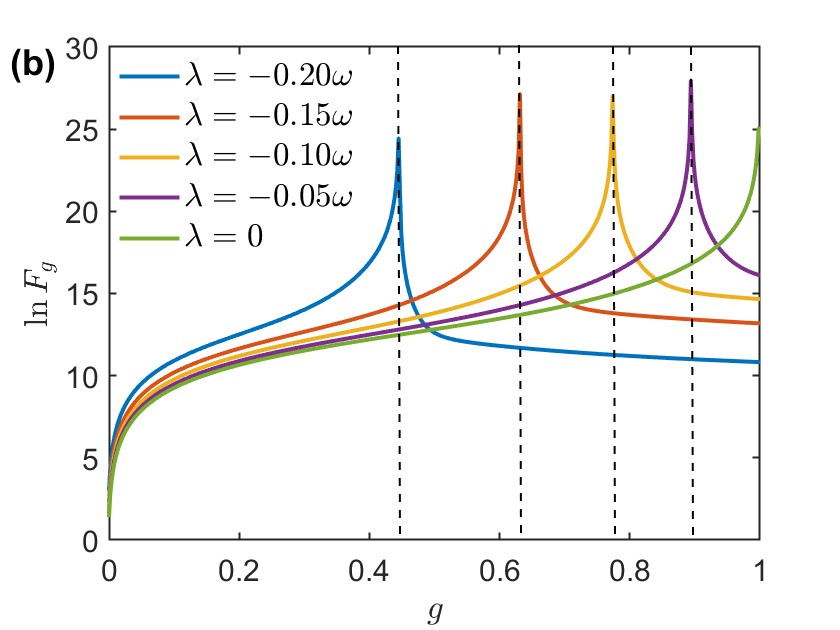}
    \includegraphics[width=7cm]{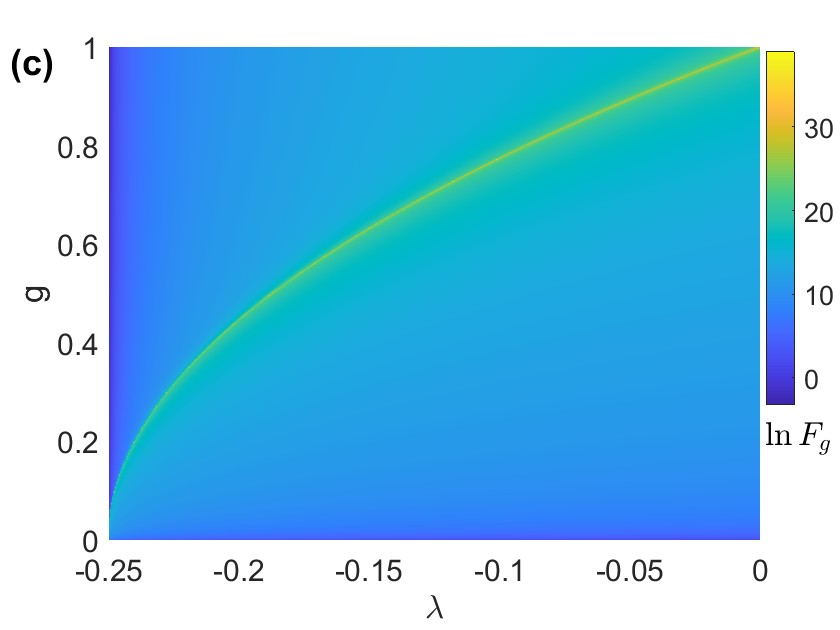}
	\caption{Globalized critical quantum metrology.
(a) Time evolution of the quantum Fisher information (QFI) $F_g(t)$ for $g = 0.097$ (lower line), $g = 0.098$ (middle line), $g = 0.099$ (upper line) in the vicinity of critical coupling point $g = 0.1$ at $\lambda =-0.2475 \omega$.
(b) Logarithm of $F_g$ as a function of $g$ at time moment $t=1000\omega$ at $\lambda=0.0,-0.05,-0.10,-0.15,-0.20$ (from right to left). The vertical dashed lines mark the critical points $g_c^{(\lambda)}$ at the corresponding values of $\lambda$.
(c) Density plot of Logarithm of $F_g$ in the $\lambda$-$g$ plane at $t=1000/\omega$.
}
\label{Fig-global-Fg}
\end{figure}

We first show the QFI enhanced by critical quantum dynamics. We take the initial state $\lvert\varphi\rangle=\lvert\downarrow\rangle_{q}\otimes\lvert\varphi\rangle_{b}$ with $\lvert\varphi\rangle_{b}=(\lvert 0 \rangle +i\lvert 1 \rangle)/\sqrt{2}$ for the illustration. Figure~\ref{Fig-global-Fg}(a) illustrates the evolution of the QFI $F_g$ for different couplings around $g_c^{(\lambda)}$ at a given value of $\lambda$,  as plotted according to the Eq.~\eqref{Eq-Fg}. Here we illustrate by $\lambda=-0.2475\omega$ which yields a critical coupling point at $g_c^{(\lambda)}=0.1$. We see that the QFI increases in the evolution with respect to time $t$, while a coupling closer to $g_c^{(\lambda)}$ has a larger QFI in the time evolution. Moreover, the enhancement becomes more dramatic as the coupling gets closer to the critical point, with an accelerated increasing ratio. This demonstrates a criticality-enhanced measurement precision in dynamical approach of quantum metrology. In fact, although we illustrate by a certain initial state $\lvert\varphi\rangle_{0}$, this conclusion of criticality-enhanced QFI in dynamics holds generally, irrespective of the selection of the initial state.

\subsection{Criticality-enhanced QFI at different $\lambda$}

The QFI enhancements by criticality can be more clearly seen in Fig.~\ref{Fig-global-Fg}(b) where the continuous dependence of $F_g$ on the coupling is illustrated in logarithm at fixed values of $\lambda$. One can see the diverging peaks of the QFI which emerge around the critical points $g_c^{(\lambda)}$. Note that, in the absence of auxiliary nonlinear term ($\lambda=0$, rightmost line), the the diverging QFI appears only round $g=1$ while the QFI decreases soon away from $g=1$, which indicates the local limitation of CQM. Now in the presence of the auxiliary nonlinear term ($\lambda\neq 0$, other lines from the rightmost) we can also have diverging QFI away from $g=1$.

\subsection{Globalized CQM in continuously tuned $\lambda$.}

The above examples of $\lambda$ illustrate the feasibility of globalized criticality-enhanced quantum metrology. Such a globalized CQM can be more panoramically observed in Fig.~\ref{Fig-global-Fg}(c) which shows the in logarithm of $F_g$ in the $g$-$\lambda$ plane. We see that, as $\lambda$ is adjusted, the diverging peak of the QFI can be continuously tuned from $g=1$ down to the $g=0$ limit, covering the entire coupling regime. Of course, a diverging QFI peak at couplings stronger than $g=1$ can also be yielded by a positive $\lambda$. Thus, we realize the globalization of the CQM in the QRM by the auxiliary nonlinear term.

\section{Measurement schemes}\label{Sect-Measurement-schemes}

The square root of the QFI is the upper bound of measurement precision in quantum metrology, as established by the quantum
Cram\'{e}r-Rao theorem~\cite{Cramer-Rao-bound}. In practice we need some physical properties in measurements. Here we examine the expectation value of quadrature operator $\left\langle X\right\rangle_t$ in the time evolution.

\subsection{Dynamical equations for $\left\langle X\right\rangle_t$}

To track the dynamics of $\left\langle X\right\rangle_t$ we solve the Heisenberg equation for Hamiltonian \eqref{Hnp}, which leads us to
\begin{equation}
	\left\langle X\right\rangle_t = \frac{\sqrt{2}}{2} \epsilon_g ^{-\frac{1}{2}}\sin(\frac{\sqrt{\epsilon}}{2}t),
\label{Eq-X}
\end{equation}
so that
\begin{equation}
\partial_g\left\langle X\right\rangle _t = \frac{\sqrt{2}\omega g}{2(\omega+4\lambda)}\epsilon_g^{-\frac{3}{2}}[\sin(\frac{\sqrt{\epsilon}}{2}t)-\frac{\sqrt{\epsilon}}{2}t \cos(\frac{\sqrt{\epsilon}}{2}t)],
\label{Eq-dXdg}
\end{equation}
In the same way, we can obtain
\begin{equation}\label{Eq-DeltaX}
	\left\langle X^2\right\rangle_t = 1+4\omega^2 g^2\epsilon^{-1} \sin^2(\frac{\sqrt{\epsilon}}{2}t).
\end{equation}
Thus, the varaiance of $\left\langle X\right\rangle_t$ is extracted to be
\begin{equation}
		\begin{aligned}
			(\Delta X)^2 = \left\langle X^2\right\rangle_t -\left\langle X\right\rangle_t^2
			=1+\left(\frac{1}{2\epsilon_g}-1\right)\sin^2(\frac{\sqrt{\epsilon}}{2}t).
		\end{aligned}
\label{Eq-X2}
\end{equation}

\subsection{Sensitivity resource extended to weak couplings}

\begin{figure}[t]
	\centering
	\includegraphics[width=7cm]{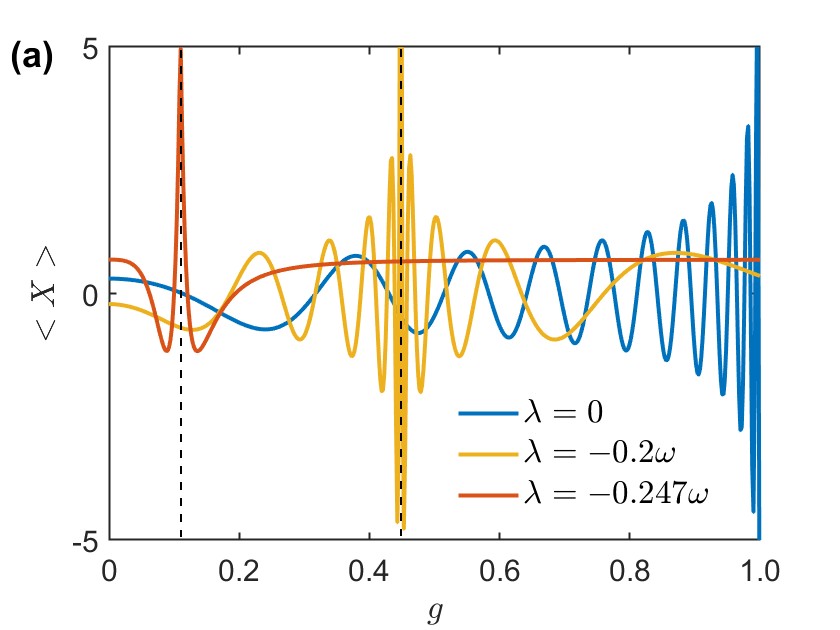}
	\includegraphics[width=7cm]{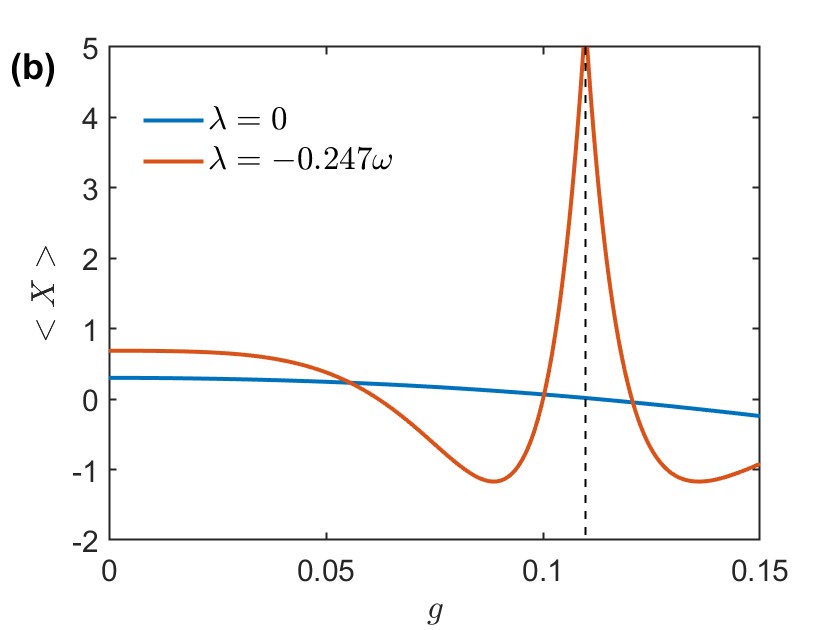}
	\caption{Measurement scheme: Sensitivity resource of $\left\langle X\right\rangle _t$ extended from $g=1$ to weak couplings. (a) Variations of $\left\langle X\right\rangle _t$ with respect to $g$ for $\lambda=0$ (rightmost), $\lambda=-0.2\omega$ (middle) and $\lambda=-0.247\omega$ (lefttmost). (b) A close-up view of $\left\langle X\right\rangle _t$ around $g=0.1$ for $\lambda=-0.247\omega$ (peaked line) $\lambda=0$ (flat line). Here $t=75 /\omega$.}
\label{Fig-X}
\end{figure}

Some examples of $\left\langle X\right\rangle_t$ are presented in Fig.~\ref{Fig-X}(a) as functions of the coupling $g$. We see that, in the absence of the auxiliary term ($\lambda =0$, rightmost), $\left\langle X\right\rangle_t$ is varying quickly around $g_c^{(0)}=1$. We can apply such a sensitivity resource for refined measurements of $g$, as $\left\langle X\right\rangle_t$ depend on $g$ via $\epsilon _g$ which is introduced in Eq.~\eqref{Hnp}. However, away from $g_c^{(0)}$ the variation of $\left\langle X\right\rangle_t$ slows down in a great deal, which would reduces the measurement precision. Nevertheless, we can turn on the auxiliary term ($\lambda \neq0$) to regain the high sensitivity in the fast variations of $\left\langle X\right\rangle_t$ away from $g_c^{(0)}=1$ (vertical dashed lines), as illustrated by $\lambda =-0.2\omega$ (middle line) and $\lambda =-0.247\omega$ (leftmost line) around $g_c^{(\lambda)}$. A close-up plot for the case of $\lambda =-0.247\omega$ is presented in Fig.~\ref{Fig-X}(b) which provides a better view for the fast variation of $\left\langle X\right\rangle_t$ around $g_c^{(\lambda)}\approx 0.11$.

Thus, manipulation of the critical behavior of $\left\langle X\right\rangle_t$ in dynamics by adjusting $\lambda$ to $\lambda _c$ in Eq.~\eqref{eq-A2C} provides a measurement scheme in practice. Such a scheme extends the CQM globally to the entire coupling regime including the weak coupling limit.

\begin{figure}[t]
	\centering
	\includegraphics[width=7cm]{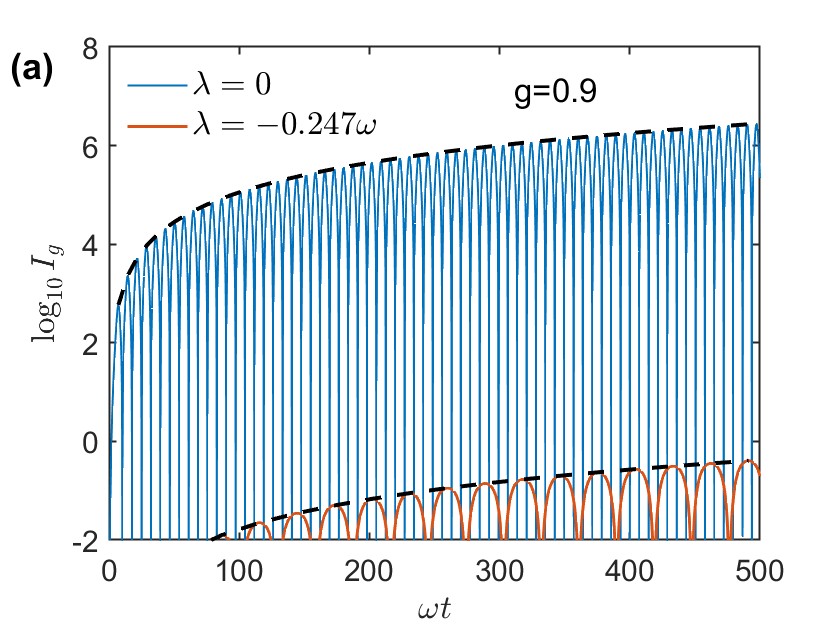}
	\includegraphics[width=7cm]{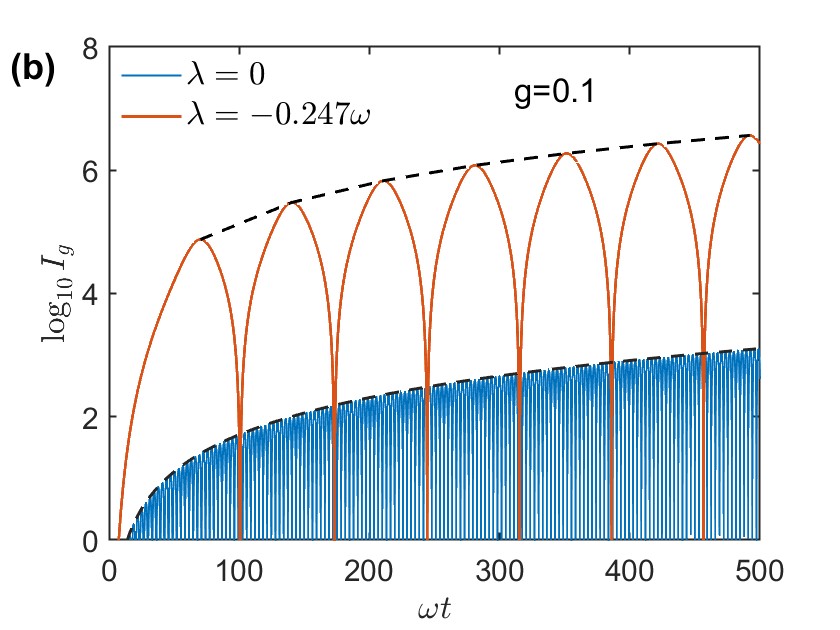}
	\caption{Globally available high measurement precision available from dynamics of $\left\langle X\right\rangle _t$. Time evolution of the inverted variance $I_g$ at $g=0.9$ (a) and at $g=0.1$ (b).  Red (blue) lines represent the $\lambda=0$ ($\lambda=-0.247\omega$) case [blue lines are higher in (a) and lower in (b)]. The dashed lines mark the peak values.}
\label{Fig-Ig}
\end{figure}

\subsection{Enhanced inverted variance for $\left\langle X\right\rangle_t$}

The accuracy of the parameter estimation $g$ via $\left\langle X\right\rangle_t$ can be characterized by the inverted variance $I_g(t)=\frac{\partial_g\left\langle X\right\rangle _t^2}{(\Delta X)^2}$~\cite{Chu2021-Metrology}. Indeed, the factor $1/(\Delta X)^2$ in $I_g(t)$ is the precision in measuring $\left\langle X\right\rangle_t$, while the numerator $(\partial_g\left\langle X\right\rangle _t)^2$ reflects the sensitivity of $\left\langle X\right\rangle _t$ in response to the coupling variations. Here we have the explicit inverted variance
\begin{equation}
		\begin{aligned}
		I_g(t)=\frac{\omega^2 g^2 [\sin(\frac{\sqrt{\epsilon}}{2}t)-\frac{\sqrt{\epsilon}}{2}t \cos(\frac{\sqrt{\epsilon}}{2}t)]^2}{(\omega+4\lambda)^2\epsilon_g^3[2+(\epsilon_g^{-1}-2)\sin^2(\frac{\sqrt{\epsilon}}{2}t)]}.
		\end{aligned}
\label{Eq-Ig} 
\end{equation}
We show the time evolution of the inverted variance $I_g(t)$ in Fig.~\ref{Fig-Ig}. One can see that $I_g(t)$ (solid lines) manifests peak profiles periodically, while the peak values (dashed lines) increases with the appearing times of the peaks. The case $\lambda =0$ exhibits high a measurement precision around $g_c^{(0)}=1$, as seen by the large values (higher blue lines) of $I_g$ in Fig.~\ref{Fig-Ig}(a) with $g=0.9$, while the measurement precision is much reduced in smaller $g$ as indicated by the smaller values (lower blue lines) of $I_g$ in Fig.~\ref{Fig-Ig}(b) with $g=0.1$. The measurement precision in smaller $g$ is restored by the auxiliary term as demonstrated by the $\lambda =-0.247\omega$ case, with the values of $I_g$ raised by four orders (higher red lines) in Fig.~\ref{Fig-Ig}(b) compared to the $\lambda =0$ case (lower blue lines).

\subsection{Scaling of QFI and inverted variance of $\left\langle X\right\rangle_t$}

\begin{figure}[t]
	\centering
	\includegraphics[width=7cm]{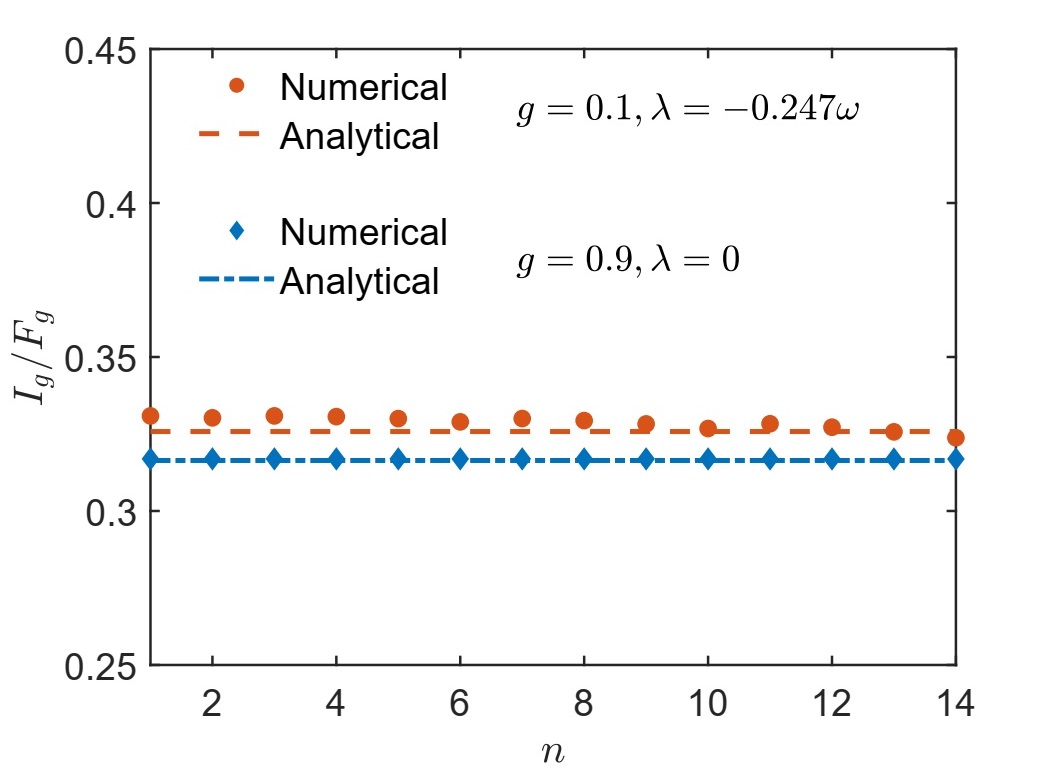}
	\caption{Scaling relation of the inverted variance ($I_g$) and the QFI ($F_g$).  The numerical data are represented by the diamonds ($\lambda=0$, $g=0.9$) and dots ($\lambda=-0.247\omega$, $g=0.1$), while the analytical results are denoted by the dash-dotted line ($\lambda=0$, $g=0.9$) and dashed line ($\lambda=-0.247\omega$, $g=0.1$). Te integer number $n$ labels the peaks of $I_g$ in time evolution.}
\label{Fig-Ratio-Ig-Fg}
\end{figure}

It turns out that the peak values of the inverted variance scale with the QFI.
Let us figure out the optimal measurement time moments and the corresponding maximum inverted variances. The periodic variations of $\partial_g\left\langle X\right\rangle _t$ and $(\Delta X)^2$ in Eqs.~\eqref{Eq-dXdg} and \eqref{Eq-X2} are driven by trigonometric functions with angular frequency $\sqrt{\epsilon}/2$, which means that the timescale for measurements is $2\pi/\sqrt{\epsilon}$. On this timescale, taking into account the condition of $\epsilon_g\rightarrow 0$ around the critical point, the two terms of $\partial_g\left\langle X\right\rangle _t$ are both proportional to $\epsilon_g ^{-3/2}$. Then the maximum of inverted variance is more determined by the minimum of $(\Delta X)^2$ in the denominator of $I_g(t)$. With this consideration, we get the optimal times,
\begin{equation}
	 \tau_n=2n\pi/\sqrt{\epsilon}
\end{equation} 
where $n$ is an integer number labeling the series of the $I_g(t)$ peaks, by the vanishing of the $\sin^2(\frac{\sqrt{\epsilon}}{2}t)$ term of $(\Delta X)^2$ in Eq.~\eqref{Eq-DeltaX}. These optimal times finally are also favorable to maximize the contribution of the second term of $\partial_g\left\langle X\right\rangle _t$ in Eq.~\eqref{Eq-dXdg} who is growing with time. Thus, the peak values of inverted variance can be extracted to be
\begin{equation}\label{equ18}
	 I_g(\tau_n)=\frac{n^2\pi^2\omega^2g^2}{2(\omega+4\lambda)^2}\epsilon_g^{-3}.
\end{equation}
At the above optimal times of $\left\langle X\right\rangle_t$, the QFI can also be calculated as
\begin{equation}\label{equ19}
    F_g(\tau_n)\simeq\frac{n^2\pi^2g^2}{\omega(\omega+4\lambda)^5}\epsilon_g^{-3}Var[N]_{\lvert\varphi\rangle}.
\end{equation}
Obviously, $I_g(\tau_n)$ and $F_g(\tau_n)$ are in the same order of magnitude, scaling with each other by a factor
\begin{equation}
		\begin{aligned}
		\frac{I_g(\tau_n)}{F_g(\tau_n)}\simeq\frac{\omega^3(\omega+4\lambda)^3}{2Var[N]_{\lvert\varphi\rangle}} 
		= \frac{1}{2Var[P^2-\epsilon_gX^2]_{\lvert\varphi\rangle}}.
		\end{aligned}
\label{Eq-Scaling}
\end{equation}
Figure~\ref{Fig-Ratio-Ig-Fg} illustrates the ratio of $I_g(\tau_n)$ and $F_g(\tau_n)$ at $g=0.9$ and $g=0.1$ with $\lambda =0$ and $g=-0.247\omega$ respectively, the scaling relation is confirmed by the agreements of the numerical results (symbols) and the analytical ones by Eq.~\eqref{Eq-Scaling} (lines).
\begin{figure}[h]
	\centering
	\includegraphics[width=8cm]{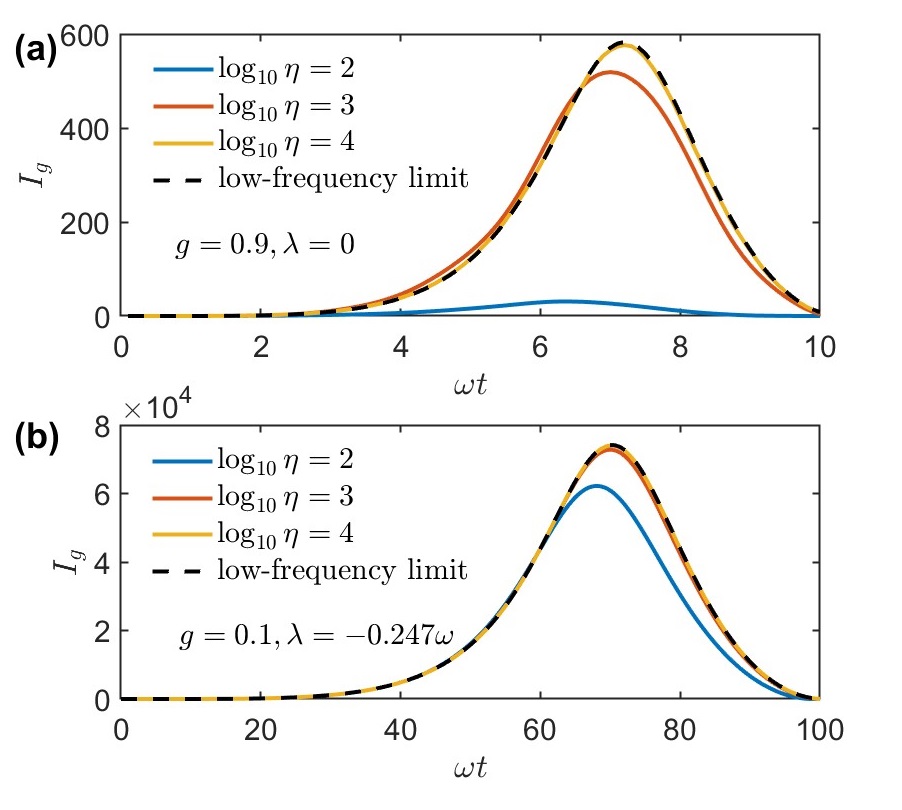}
	\caption{Surviving precision resources at finite frequencies.
(a) Inverted variance $I_g(t)$ at $\lambda =0, g=0.9$ for different frequency ration $\eta$.
(b) Inverted variance $I_g(t)$ at $\lambda =-0.247\omega, g=0.1$ for different frequency ration $\eta$.}
\label{Fig-Ig-finite-w}
\end{figure}

\section{Discrepancy of analytical results and scaling relation at finite Frequencies}\label{Sect-Scaling-relation}

The analytical results in previous sections are extracted in the low-frequency limit $\omega/\Omega \rightarrow 0$. At finite frequencies it is difficult to obtain analytical results. Nevertheless, we can simply estimate the order of the discrepancy from the results in low-frequency limit. A finite frequency would add a correction in the leading order $\mathcal{O}(\omega\eta^{-1})$ of the inverse frequency ratio $\eta=\Omega/\omega$ to the Schrieffer-Wolff transform $H_{S}=H_{np}+\mathcal{O}(\omega\eta^{-1})$ where $\omega$ is the unit. On the other hand, the quadrature operator in the transform has a higher order of correction $e^{-S}Xe^{S}=X+\mathcal{O}(\eta^{-\frac{3}{2}}).$ Thus, it can be proved that the expectation evolution of the operator $X$ at finite frequency, $\left\langle X\right\rangle _t^{\omega}$, has a leading order correction of $\mathcal{O}(\eta^{-1})$ relatively to the expression of $\left\langle X\right\rangle _t$ analytically extracted in the low-frequency limit:
\begin{equation}\label{equ20}
	\begin{aligned}
		\left\langle X\right\rangle _t^{\omega} &=\langle\varphi\lvert e^{S}e^{iH_{S}t}e^{-S}Xe^{S}e^{-iH_{S}t}e^{-S} \lvert\varphi\rangle\\
		&=\left\langle X\right\rangle _t+\mathcal{O}(\eta^{-1}),
	\end{aligned}
\end{equation}
and similarly
\begin{equation}\label{equ21}
	\begin{aligned}
		\left\langle X^2\right\rangle _t^{\omega}& =\langle\varphi\lvert e^{S}e^{iH_{S}t}e^{-S}X^2e^{S}e^{-iH_{S}t}e^{-S} \lvert\varphi\rangle\\
		&=\left\langle X^2\right\rangle _t+\mathcal{O}(\eta^{-1}).
	\end{aligned}
\end{equation}
Consequently, the discrepancy of the inverted variance in the vicinity of the critical point is approximately proportional to the frequency ratio
\begin{equation}
	\begin{aligned}
	\delta=\frac{I_g^{\omega}(\tau_n)-I_g(\tau_n)}{I_g(\tau_n)}\sim \eta^{-1}.
	\end{aligned}
\label{Eq-error-scaling-w}
\end{equation}

The time evolutions of the inverted variance $I_g$ at different frequencies are illustrated in Fig.~\ref{Fig-Ig-finite-w}(a) for $g =0.9$ with $\lambda =0$ and Fig.~\ref{Fig-Ig-finite-w}(b) for $g =0.1$ with $\lambda =-0.247\omega$. We see that the finite frequencies (solid lines) lead to some reductions of the inverted variance from the low-frequency limit (dashed line). Nevertheless, it does not downgrade much the order of inverted variance. This indicates the surviving precision resources at finite frequencies. Indeed, up to $\eta\sim 10^3$, the finite frequencies can achieve a measurement precision similar to that of the low-frequency limit.

\begin{figure}[h]
	\centering
	\includegraphics[width=7cm]{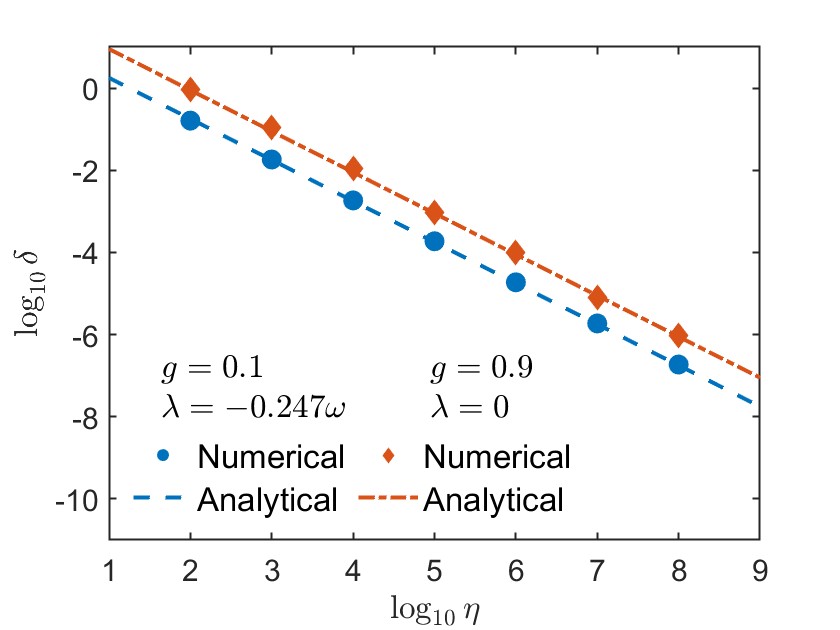}
	\caption{Scaling relation for the discrepancy of the inverted variance ($\delta$) and the frequency ratio ($\eta =\Omega/\omega$). The numerical data are represented by the diamonds ($\lambda=0$, $g=0.9$) and dots ($\lambda=-0.247\omega$, $g=0.1$), while the analytical results are denoted by the dash-dotted line ($\lambda=0$, $ g=0.9$) and dashed line ($\lambda=-0.247\omega$, $g=0.1$).}
\label{Fig-error-scaling-w}
\end{figure}

The scaling relation in Eq.~\eqref{Eq-error-scaling-w} for the discrepancy at finite frequencies is confirmed by Fig.~\ref{Fig-error-scaling-w} where we see the agreements of the numerical data (symbols) and the analytical results for both cases of $ g=0.9,\lambda =0$ and $g=0.1,\lambda =-0.247\omega$. From the scaling relation we find that the auxiliary term ($\lambda \neq0$) not only globalizes the CQM but also reduces the discrepancy of the inverted variance at finite frequencies compared to the original QRM ($\lambda = 0$).

\begin{figure}[t]
	\centering
	\includegraphics[width=7.5cm]{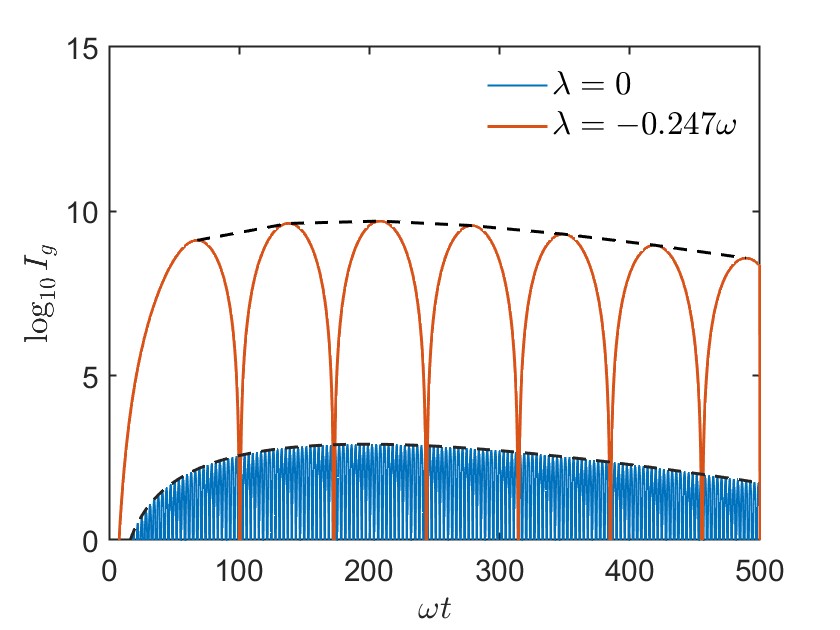}
	\caption{Surviving precision resources in the decoherence. Here $\lambda=-0.247\omega$, $g=0.1$.}
\label{Fig-decoherence}
\end{figure}

\section{In the presence of decoherence}\label{Sect-decoherence}

To include the environmental influence in a more realistic situation, we further consider the presence of decoherence~\cite{Hwang2018DissipativeRabi,AiQ2023,Vaaranta2025JCRabiLindblad}. The dissipative dynamics is governed by the Lindblad master equation
\begin{equation}\label{equ23}
	\dot{\rho}=-i\left[ H_{np},\rho \right] +\gamma_a D[a]+\gamma_h D[a^\dagger] ,
\end{equation}
where $\rho$ is the density matrix, $D[a]=a\rho a^\dagger-a^\dagger a\rho /2-\rho a^\dagger a /2$ denotes the dissipation superoperators, and $\gamma_a$,$\gamma_h$ are the decay rate and heating rate. The dynamical equations for the expectation values can be obtained as
	\begin{equation}\label{equ24}
	\begin{aligned}
		&\partial_t{\left\langle X\right\rangle }= \bar{\omega}\left\langle P\right\rangle - \frac{\gamma_-}{2}\left\langle X\right\rangle ,\\
		&\partial_t{\left\langle P\right\rangle} = -\frac{\epsilon}{4\bar{\omega}}\left\langle X\right\rangle - \frac{\gamma_-}{2}\left\langle P\right\rangle ,\\
		&\partial_t{\left\langle X^2\right\rangle} =
		-\gamma_-\left\langle X^2\right\rangle +\bar{\omega}\left\langle G\right\rangle +\frac{\gamma_+}{2} ,\\
		&\partial_t{\left\langle P^2\right\rangle } =
		-\gamma_-\left\langle P^2\right\rangle -\frac{\epsilon}{4\bar{\omega}}\left\langle G\right\rangle +\frac{\gamma_+}{2} ,\\
		&\partial_t{\left\langle G\right\rangle }= 	-\gamma_-\left\langle G\right\rangle  +2\bar{\omega}\left\langle P^2\right\rangle  -\frac{\epsilon}{2\bar{\omega}}\left\langle X^2\right\rangle ,
	\end{aligned}		
\end{equation}
where $G=XP+PX$, $\bar{\omega}=\sqrt{\omega^2+4\lambda\omega}$, $\gamma_+=\gamma_a+\gamma_h$, $\gamma_-=\gamma_a-\gamma_h$ and the expectation value of operator $O$ is defined as $\langle O \rangle={\rm tr}[ O \rho]$.

With the initial state $\lvert\varphi\rangle$ the above dynamical equations can be solved with explicit results for $\left\langle X\right\rangle _t$
\begin{equation}
		\begin{aligned}
			&\left\langle X\right\rangle _t = \frac{\sqrt{2}}{2}\epsilon^{-\frac{1}{2}} \sin(\frac{\sqrt{\epsilon}}{2}t)e^{-\frac{1}{2}\gamma_-t},\\
			&\partial_g\left\langle X\right\rangle _t = \frac{\sqrt{2}\omega g \  \epsilon_g^{-\frac{3}{2}} } {2(\omega+4\lambda)} [\sin(\frac{\sqrt{\epsilon}}{2}t)-\frac{\sqrt{\epsilon}}{2}t \cos(\frac{\sqrt{\epsilon}}{2}t)]e^{-\frac{\gamma_-t}{2}}
		\end{aligned}
\label{Eq-X-decoherence}
\end{equation}
Furthermore, the equations for the variances can be constructed as
\begin{equation}\label{equ26}
	\begin{aligned}
		&\partial_t(\Delta X)^2 =
		-\gamma_-(\Delta X)^2 +\bar{\omega}\tilde{G} +\frac{\gamma_+}{2} ,\\
		&\partial_t(\Delta P)^2 =
		-\gamma_-(\Delta P)^2 -\frac{\epsilon}{4\bar{\omega}}\tilde{G} +\frac{\gamma_+}{2} ,\\
		&\partial_t \tilde{G}= 	-\gamma_-\tilde{G}  +2\bar{\omega}(\Delta P)^2 -\frac{\epsilon}{2\bar{\omega}}(\Delta X)^2,
	\end{aligned}		
\end{equation}
where $\tilde{G}=\left\langle G\right\rangle-2\left\langle X\right\rangle\left\langle P\right\rangle$, which lead us to the solution
\begin{equation}
		\begin{aligned}
(\Delta X)^2 =& \frac{1}{4}\left\{2+\frac{1}{\epsilon_g}+ \frac{\gamma_-\gamma_+(\epsilon-4\bar{\omega}^2)}{\epsilon(\gamma_-^2+\epsilon)} \right.\\
			&\left.+ \frac{\gamma_+(2\gamma_-^2+\epsilon+4\bar{\omega}^2)}{\gamma_-(\gamma_-^2+\epsilon)}(e^{\gamma_-t}-1) \right.\\
			&\left. + \left[2-\frac{1}{\epsilon_g}- \frac{(\epsilon-4\bar{\omega}^2)\gamma_+\gamma_-}{\epsilon(\gamma_-^2+\epsilon)}\right] \cos(\sqrt{\epsilon}t) \right.\\
			&\left. - \frac{4\omega^2 g^2\gamma_+}{\sqrt{\epsilon}(\gamma_-^2+\epsilon)} \sin(\sqrt{\epsilon}t) \right\}e^{-\gamma_-t}
		\end{aligned}		
	\end{equation}
\label{Eq-X2-decoherence}
Then the inverted variance in the presence of decoherence can be obtained as
\begin{equation}
		\begin{aligned}
	I_g(t) =&\frac{\omega^2g^2}{(\omega+4\lambda)^2\epsilon_g^3}\left[\sin(\frac{\sqrt{\epsilon}}{2}t)-\frac{\sqrt{\epsilon}}{2}t\cos(\frac{\sqrt{\epsilon}}{2}t)\right]^2 \\
			&\left. \middle/ \left\{2+\frac{1}{\epsilon_g}+ \frac{\gamma_-\gamma_+(\epsilon-4\bar{\omega}^2)}{\epsilon(\gamma_-^2+\epsilon)} \right.\right.\\
			&\left.+ \frac{\gamma_+(2\gamma_-^2+\epsilon+4\bar{\omega}^2)}{\gamma_-(\gamma_-^2+\epsilon)}(e^{\gamma_-t}-1) \right.\\
			&\left. + \left[2-\frac{1}{\epsilon_g}- \frac{(\epsilon-4\bar{\omega}^2)\gamma_+\gamma_-}{\epsilon(\gamma_-^2+\epsilon)}\right] \cos(\sqrt{\epsilon}t)\right.\\
			&\left. - \frac{4\omega^2 g^2\gamma_+}{\sqrt{\epsilon}(\gamma_-^2+\epsilon)} \sin(\sqrt{\epsilon}t) \right\}   	
		\end{aligned}
\label{Eq-Ig-decoherence}
\end{equation}
Setting $\gamma_\pm=0$ in Eqs.~\eqref{Eq-X-decoherence}, \eqref{Eq-X2-decoherence} and \eqref{Eq-Ig-decoherence} will retrieve $\left\langle X\right\rangle _t$  $\partial_g\left\langle X\right\rangle _t$, $(\Delta X)^2 $,  and $I_g(t)$ in Eqs.~\eqref{Eq-X}, \eqref{Eq-dXdg}, \eqref{Eq-X2} and \eqref{Eq-Ig} previously obtained in the absence of decoherence.

Taking $\gamma_-=0.01\omega$ and $\gamma_+=0.03\omega$ as an illustration, the dynamics of $I_g(t)$ in decoherence is shown in Fig.~\ref{Fig-decoherence}. We see that the main features in the dynamics remain despite the tendency that after a rising the peaks of $I_g(t)$ tend to get lower in the long-time limit. Nevertheless, the lost high values of the inverted variance for $\lambda =0$ in weak couplings ($g=0.1$ illustrated here in Fig.~\ref{Fig-decoherence}) is regained by turning on the auxiliary term (here $\lambda=-0.247$).
Thus, we see that the advantage of globalization of CQM for the QRM by the auxiliary term is still retained in the presence of decoherence.

\section{Conclusions}\label{Section-Conclusion}

We have proposed a globalized CQM in the QRM by adding an auxiliary nonlinear term which can extend the critical point to a continuous critical regime. We have analytically analyzed the critical quantum dynamics of the QFI which not only manifests a divergently high value and but also is globally available by tuning the auxiliary term.
As a result, a high measurement precision is globally accessible over the entire coupling regime from the original critical point of the QRM down to the weak-coupling limit. This breaks the local limitation of the CQM in the QRM and globalizes this paradigmatic CQM.

We have further shown a measurement scheme to realize the globalized CQM by the critical quantum dynamics of the quadrature expectation $\left\langle X\right\rangle _t$. In fact, $\left\langle X\right\rangle _t$ varies quickly around the continuously tunable critical coupling point, which provides the critical sensitivity resource for the quantum measurements. Indeed, the inverted variance exhibits high peaks in dynamics which characterize the high measurement precision. In the QRM such a quick variation of $\left\langle X\right\rangle _t$ lies around the scaled coupling $g=1$ but slows down away, indicating the loss of high measurement precision. Nevertheless the quick variation of $\left\langle X\right\rangle _t$ and high measurement precision are restored by the the auxiliary term. We have figured out the optimal measurement times and extracted the peak values of the inverted variance in the presence of the auxiliary term. We have demonstrated that the inverted variance scales with the QFI, which implies the feasibility of the globalized CQM.

In addition, we have estimated the discrepancy between the finite frequencies and the low-frequency limit. We find that the order of measurement precision is retained up to a considerable finite frequency regime. We have also revealed the scaling relation for the discrepancy and the frequency, which provides a convenient reference for the precision estimation at finite frequencies.

We finally address the influence of dissipation by solving the Lindblad master equation. We have found the dynamical features of the inverted variance remain in the presence of decoherence and the advantage of globalization of CQM for the QRM by the auxiliary term is still retained.

As a closing remark, the the auxiliary term for the QRM can be realized and adjustable in hybrid systems with optomechanical coupling~\cite{Chen2021A2NC,Lv2018A2PRA,Lv2018A2PhysRevApplied} or Kerr magnons~\cite{LiuGang2023}. Thus, our proposal for globalization of CQM for the QRM may provide a feasible protocol.

\begin{acknowledgments}
This work was supported by the National Natural Science Foundation of China
(Grants No. 12474358, No. 11974151, and No. 12247101).
\end{acknowledgments}

\appendix\bigskip

\section{Formulation for the coupling regime beyond the critical point ($g > g_c ^{\lambda}$)}\label{Appendix-Beyond-gc}

In the QRM, the analytical results beyond the QPT is different from the regime before the QPT~\cite{Hwang2015PRL}. Similarly, in the presence of the auxiliary term, for the coupling regime after the critical point, i.e. $g > g_c ^{\lambda}$, the analytical results need to be adjusted, as addressed here in this Appendix. 

Indeed, since both squeezing and displacement occur after the QPT~\cite{Ying2015}, a displacement transform is also needed here for the $g > g_c ^{\lambda}$ regime:
\begin{equation}\label{A1}
\begin{aligned}
	H_{\alpha} & =\mathcal{D}^\dagger(\alpha)H_{r}\mathcal{D}(\alpha) \\
		& =\omega \left( 1+\frac{4\lambda}{\omega}\right) ^\frac{1}{2}(a^\dag+\alpha) (a+\alpha)+\frac{\Omega}{2}\sigma_{z}\\
		&\quad+\frac{\sqrt{\omega\Omega}}{2}g\left( 1+\frac{4\lambda}{\omega}\right) ^{-\frac{1}{4}}(a+a^\dag+2\alpha)\sigma_{x},
\end{aligned}
\end{equation}
where $\mathcal{D}(\alpha)=e^{\alpha (a^\dagger - a)}$ is the displacement operator, with $4\alpha^2=\Omega g^{-2} (\omega^2+4\lambda\omega)^{\frac{3}{2}} [\omega^2 g^4 - (\omega+4\lambda)^2]$. 

Let us define the rotated spin basis $\widetilde{\lvert\downarrow\rangle} = \cos{\theta} \lvert\uparrow\rangle + \sin{\theta} \lvert\downarrow\rangle$ and $\widetilde{\lvert\uparrow\rangle} = -\sin{\theta} \lvert\uparrow\rangle + \cos{\theta} \lvert\downarrow\rangle$. To eliminate the $\alpha \sigma _x$ term, the spin rotation angle $\theta$ is determined as $\tan{2\theta} = 2g\alpha\sqrt{\omega/\Omega}\left( 1+4\lambda/\omega\right) ^{-\frac{1}{4}}$. In terms of Pauli matrices $\{\tilde{\sigma}_{x}, \tilde{\sigma}_{y}, \tilde{\sigma}_{z}\}$ on the new
basis $\{\widetilde{\lvert\downarrow\rangle}, \widetilde{\lvert\uparrow\rangle}\}$, Eq.~\eqref{A1} becomes
\begin{equation}\label{A2}
	\begin{aligned}
		H_{\alpha} &
		=\omega \left( 1+\frac{4\lambda}{\omega}\right) ^\frac{1}{2}(a^\dag a + \alpha^2) +\frac{\Omega_\alpha}{2}\tilde{\sigma}_{z}\\
		&\quad+\frac{\sqrt{\omega\Omega_\alpha}}{2}g_\alpha\left( 1+\frac{4\lambda}{\omega}\right) ^{-\frac{1}{4}}(a+a^\dag)\tilde{\sigma}_{x}.
	\end{aligned}
\end{equation}
where $\Omega_\alpha=\frac{\Omega\omega g^2}{\omega+4\lambda}$ and $g_\alpha=g^{-2}(1+\frac{4\lambda}{\omega })^{\frac{3}{2}}$. 

Now that $H_{\alpha}$ has the similar form to $H_{r}$ in the $g < g_c ^{\lambda}$ regime, using the same methods in the main text the effective low-energy Hamiltonian, the QFI and the observable can be calculated to be
\begin{equation}\label{A3}
	H_{np}^\alpha=\frac{\sqrt{\omega^2+4\lambda\omega}}{2}\left[P^2+\epsilon_g^\alpha X^2\right] ,
\end{equation}
\begin{equation}\label{A4}
	\begin{aligned}
		F^\alpha_g\simeq 64(\frac{1-\epsilon_g^\alpha}{g})^2\frac{[\sin(\sqrt{\epsilon_\alpha} t)-\sqrt{\epsilon_\alpha} t]^2}{\epsilon_\alpha^3}Var[N_\alpha]_{\lvert\varphi\rangle}
	\end{aligned}.
\end{equation}
\begin{equation}\label{A5}
	\left\langle X\right\rangle_t^\alpha = \sqrt{\frac{1}{2\epsilon_g^\alpha}}  \sin(\frac{\sqrt{\epsilon_\alpha}}{2}t),
\end{equation}
where  $\epsilon_g^\alpha=\left(1-(\frac{\omega+4\lambda}{\omega g^2})^2\right)$,  $\epsilon_\alpha=4\omega(\omega+4\lambda)\epsilon_g^\alpha$, and $N_\alpha=(\omega^2+4\lambda\omega)^\frac{3}{2}(P^2-\epsilon_g^\alpha X^2)$.

\bibliography{Refs-2025-g1-A2-Cover-refs}

\end{document}